\documentclass[12pt,a4paper]{article}
\usepackage{graphicx,amssymb}
\begin{document}

\title{
\begin{flushright}
\normalsize ULB-TH/04-29
\end{flushright}
\vskip 10mm
A comparative study of correlations between arrival directions of
ultra-high-energy cosmic rays and positions of their potential
astrophysical sources}
\author{D.\,S.\,Gorbunov$^{1}$, S.\,V.\,Troitsky$^{1,2}$\\
\small\it $^{1}$~Institute for Nuclear Research of the Russian Academy of
Sciences,\\
\small\it
60th October Anniversary Prospect 7a, 117312, Moscow, Russia\\
\small\it
$^{2}$~Service de Physique Th\'{e}orique, CP 225,\\
\small\it
  Universit\'{e} Libre de Bruxelles, B--1050, Brussels, Belgium}
\date{}
\maketitle
\begin{abstract}
{We consider various classes of persistent extragalactic astrophysical
sources which have been suggested in literature as possible emitters
of ultra-high-energy cosmic rays (UHECR).
We compare the strength of the claimed
correlations by a uniform procedure for all classes of sources by
making use of the AGASA, Yakutsk and HiRes stereo data. BL Lac type
objects correlate with the cosmic rays detected by all three
independent experiments and are more probably, compared to other
astrophysical sources, related to the UHECR origin. With the
account of the Galactic magnetic field (not possible for the HiRes
data at the moment), apart of BL Lac type objects, unidentified
gamma-ray sources may be
correlated with AGASA and Yakutsk cosmic rays. }
\end{abstract}

\section{Introduction} Thousands of cosmic rays with energies
higher than $10^{19}$~eV have been detected by various experiments. Still,
the origin of these energetic particles remains unknown. Deflections of
charged particles by cosmic magnetic fields are relatively small (though
significant) at these energies; moreover, presence of a fraction of
neutral particles, which propagate rectilinearly, is not excluded. This
opens a possibility for direct searches of astrophysical sources of the
high-energy cosmic rays by positional correlations.

Several difficulties limit the application of this approach. Firstly, at
least a large fraction of air showers with $E\gtrsim 10^{19}$~eV is
believed to be caused by protons, whose trajectories are bend by
cosmic magnetic fields by several degrees. Lack of
knowledge about these fields makes it difficult to account for the
deflections while at highest energies, where the deflections are
believed to be small, low statistics of the cosmic rays prevents one from
making definite conclusions.

Secondly, the angular resolution of the cosmic-ray experiments is very
poor by the astronomical standards. The most precise ground array, AGASA,
has an average error of $1.8^\circ$ in determination of the arrival
directions at the highest observed energies, and $2.5^\circ$ at $E\sim
10^{19}$~eV~\cite{AGASA:list}. As a result, an impressive number of
astrophysical objects fall in the error box of each particular cosmic ray.
The situation improved with the availability of the first set of the HiRes
stereo data~\cite{HiRes} with claimed angular resolution of $0.6^\circ$.
Still, this resolution is quite poor for direct identification: dozens of
optical, radio and X-ray sources are contained in a given circle of
$0.6^\circ$ radius on the Celestial sphere.

With these complications, statistical methods give the only possible
clue to search for positional correlations between the cosmic rays and
their potential sources. An excess of cosmic rays at small angular
distances (in particular, at those compared to the experiment's
angular resolution) from objects of a certain class may suggest that
this class of objects contains more probably the sources of these
cosmic rays than other classes, which do not exhibit such an
excess.\footnote{The background of random coincidences is always
present, and one cannot claim that one particular object is a source
of a given cosmic ray based on these statistical arguments only.}

Many astrophysical sites have been suggested which can host an accelerator
powerful enough to produce particles with $E\sim 10^{19}$~eV and higher.
Most of them are extragalactic (which is consistent with the lack of
global galactic anisotropy of the arrival directions of high-energy cosmic
rays). Various cosmic accelerators may operate
by similar mechanisms, so the most efficient way to distinguish between
them is based on direct positional correlations. Up to now, the
correlations with different classes of objects were studied by different
methods, with different data sets, or were not studied at all.
This work aims at filling this gap and
calculating the chance probabilities of the claimed correlations
by a uniform method, which
allows to compare the strength of correlations and to figure out more
probable sources among those
suggested previously.

The rest of the paper is organized as follows.  In Sec.~2, we discuss
the logic of the statistical analysis of correlations, describe the
methods we use and emphasize several subtle points. Sec.~3 lists
potential astrophysical sources, briefly recalls motivations for these
candidates and previous correlation studies. For each class of
sources, we give here references for the catalogs we use in the
current study. In Sec.~4, we present the results of the correlation
studies for several representative samples of cosmic rays. The
conjectures about positional correlations were formulated on the basis
of the cosmic rays observed by AGASA, and sometimes by the Yakutsk
experiment. We use the samples of cosmic rays with energies $4\cdot
10^{19}<E<10^{20}$~eV observed by AGASA (49 rays) and by AGASA and Yakutsk
(59 rays) to treat the suggested sources uniformly and to reformulate the
conjectures. {We consider independently neutral and proton primaries and
in the latter case, we account for proton deflection in the Galactic
magnetic field. } Then, we test the same conjectures with recently
published HiRes stereoscopic data (271 cosmic ray with $E>10^{19}$~eV),
assuming only neutral primaries, since energies of the HiRes events are
unpublished. Finally, we look for correlations of suggested sources with
the cosmic rays of the highest ($E>10^{20}$~eV) energies. Sec.~5 contains
conclusions and discussion.

\section{The correlation analysis}
The general logic of the correlation studies by statistical methods is
to test the hypothesis that the distribution of the observed arrival
directions of the cosmic rays is {\em isotropic}, and that there is no
excess of the cosmic rays at small angular distances\footnote{{That is, at
those of order of the experimental angular resolution.}} from the sources
listed in a given catalog (and {\em not} to test the hypothesis that the
sources from the list emit cosmic rays). To this end, a large number of
cosmic-ray events are simulated which arrive at the Earth isotropically.
Some part of them is "accepted", following the same rules which determine
the acceptance of a given cosmic-ray experiment. The angular distances
between the simulated events and the astrophysical objects from the
catalog are calculated and their distribution is compared to the
distribution of distances between the real cosmic rays and the same
objects. The less the probability that the real distribution is a
statistical fluctuation of one calculated for isotropic cosmic rays, the
stronger the indication that the objects from that particular list may be
the cosmic-ray sources.

Another way of reasoning might be to simulate the positions of the
sources and to compare them with real cosmic rays, or to simulate
both. However, the distribution of the objects of a particular class
in the Universe follows specific laws and is not easy to be
simulated. The selection biases of discovery, observation, and
inclusion in catalogs add more to this problem. On the other hand, the
directional acceptance of the cosmic-ray detectors is quite well
understood and the distribution of the observed cosmic rays agrees
with the theoretical acceptance reasonably well. That is why we
simulate the cosmic rays and not the sources.

To compare the real and simulated distributions of angular distances,
we follow a conventional procedure used in correlation studies.  One
takes a catalog of $n_r$ cosmic rays observed by a given experiment
and calculates the number $\nu (\delta )$ of events falling within the
angular distance $\delta $ from the suggested sources. The value of
$\delta $ should be of order of the experiment's angular
resolution. The actual value at which the real effect is
better seen may be determined by simulations described in
Ref.~\cite{BL1} for AGASA and Yakutsk and in Ref.~\cite{BL-HiRes} for
HiRes stereo. In all cases, the simulation gives the value of $\delta
$ somewhat larger than the angular resolution. The reason for this
smoothing is quite clear: a significant part of the observed cosmic rays
is not related to known sources included in catalogs. The Monte Carlo code
generates then a random direction in the visible half of the Celestial
sphere, determined by the azimuth angle $A$ and the zenith angle $Z$, as
well as the sidereal time of the event arrival, $s$. Then the code either
accepts (with the probability $P(A,Z,s)$) or rejects the event. The
procedure is repeated until $N$ sets, each of $n_r$ accepted events, are
generated, where $N$ is large: we use $N=10^4$ sets in our study except
for the cases of a very strong signal when we use $N=10^5$. For the ground
arrays, the probability $P(A,Z,s)=\cos Z$; for the HiRes experiment in
stereo mode, $P(A,Z,s)=P_A(A)P_Z(Z)P_s(s)$ and the functions
$P_A$,$P_Z$,$P_s$ are given in Ref.~\cite{HiRes-exposure}. For each mock
set $i=1,\dots N$, the quantity $\nu _i(\delta )$ is calculated. The
probability that the observed distribution of cosmic rays with respect to
the sources is a fluctuation of a random distribution is estimated (for
each particular pair of samples of cosmic rays and astrophysical objects)
simply by
$$
{\cal P}={ \mbox{number of mock sets with} \,
\nu _i(\delta )\ge \nu
(\delta )
\over N}.
$$
It is worth noting that the value of ${\cal P}$ should not
be, in general, interpreted as the statistically correct probability
if subsequent tries with different catalogs were made (see
Sec.~\ref{sec:tries}) and if the purpose of a study is to test the
null-hypothesis that there are no positional correlations between
cosmic rays and any of the astronomical catalogs considered.

For the
cosmic-ray experiments with relatively good angular resolution, and
assuming charged primaries, the signal may be smeared out by
deflections in the magnetic fields. One may attempt to correct the
arrival directions for bending in the Galactic magnetic field (GMF),
assuming a particular GMF model. In this study, when applicable, we
assume the GMF model used in Ref.~\cite{BL:GMF} and proton-like primaries
(electric charge $+1$). The knowledge of particle energies is crucial
for the correction for GMF and $30\%$ errors in energy determination may
result in additional increasing of the value of $\delta $. All
simulated events are subject to the same correction as the real ones;
to this end, the MC code assigns randomly the energies of $n_r$
observed events to $n_r$ simulated events in each of $N$ mock sets.

The cuts imposed by hand on the samples of cosmic rays and possible
sources may result in essential lowering of the probability ${\cal P}$. One
may impose these cuts a posteriori, to maximize the signal, and to
compensate the probability by a penalty factor (see, for instance,
Ref.~\cite{BL1}).
In this study, we do not impose any {\it additional} cuts
on the catalogs taken from original works.
All cuts in astronomical catalogs (e.g.\ the $z<0.01$ cut in a
sample of Seyfert galaxies or several cuts in a sample of dead QSO
candidates, see Sec.~3) were made {\it in the original works}
where these candidate objects had been
suggested. We refer the reader to these works for motivation of the cuts.
In the catalogs of cosmic rays, the cuts in energies and zenith angles
were made a priory as well and were determined by the availability of
data. We note here that making cuts which maximize the signal is very
useful for further detalization of the class of cosmic-ray emitters but
postpone this subject to subsequent studies.

One subtle point in this kind of a study concerns the value of the bin
size $\delta $. The probability ${\cal P}$ depends strongly on $\delta $
(see plots in Sec.~4). Hence, choice of $\delta $ may be considered as a
cut which requires its own penalty factor. However, varying $\delta $ at
scales smaller than the angular resolution of a detector, one can easily
obtain strong unphysical correlations: one single coincidence within
$0.1^\circ$ of a source and a cosmic ray observed by a detector with the
angular resolution of $2^\circ$ results in a drop of ${\cal P}$ of order
$(0.1/2)^2=2.5 \cdot 10^{-3}$ compared to one coincidence within
$2^\circ$, while physically the two cases have the same meaning (see
Fig.~3 in Sec.~4.4). Account of the correlations of this kind in simulated
sets may result in strong overestimation of the penalty factor if one
chooses to determine $\delta $ a posteriori to maximize the signal and
then to compensate for this choice by introducing penalties. In this
study, we fix $\delta $ a priori following the simulations of
Refs.~\cite{BL1,BL-HiRes}. These values are of order of the experiment's
angular resolution: $\approx 2.5^\circ$ for AGASA ($2.8^\circ$ for AGASA
with GMF correction)\footnote{We use the same values for the
AGASA-dominated joint samples of cosmic rays.} and $\approx 0.8^\circ$ for
HiRes stereo. We will see that in the cases when a strong signal is seen,
the minimal ${\cal P}$ often corresponds to a different but close value of
$\delta $.

Since we are interested in a comparative study of various astrophysical
sources only, hence in {\em relative} values of probabilities, and not in
testing a hypothesis that the objects of a particular class emit cosmic
rays, this approach is sufficient for our purposes. However, a more
adequate method to obtain {\em absolute } values of probabilities should
avoid the choice of $\delta $ at all. Indeed, there exist statistical
methods to compare distributions without use of explicit binning. Probably
the best known one is the Kolmogorov-Smirnov (KS) test; however, in our
case it is not applicable for two reasons. Firstly, the KS test does not
account for the value of experimental errors. Secondly, its sensitivity to
differences between distributions close to the end points of the interval
at which the distributions are determined is very poor, while the
positional correlations appear at values of the angular distances compared
to the detector's resolution and much less than the maximal possible
angular distance of $180^\circ$, that is very close to the zero end point
of the interval $[0^\circ,180^\circ]$. One may attempt to cut the interval
at some angle $\Delta$ in order to shift the effect towards the center of
the interval; we checked that the result is very sensitive to $\Delta $ in
this case. Methods other than the KS test exist which are more sensitive
at the ends of an interval, but, as, for instance, the Anderson-Darling
method, they are directly implemented only for the simplest distributions.
As a result, this approach awaits further studies.

\section{Potential sources}
Various classes of potential astrophysical accelerators are discussed, for
instance, in Ref.~\cite{Anch-review}. We tried to include in this study
all sources mentioned in the literature as candidate sources of
ultra-high-energy cosmic rays, subject to two criteria: we require the
objects to be persistent (gamma-ray bursts are excluded in this way) since
the comparison of significance can hardly be performed between persistent
sources and flares, and we consider only extragalactic objects since no
galactic anisotropy is seen with the impressive statistics of events at
$E>10^{19}$~eV. For the case of {\em unidentified} gamma-ray sources, to
safely exclude Galactic populations, we require the galactic latitude
$|b|>10^\circ$. For comparison, we present also results without this cut.

\subsection{Flat spectrum radio quasars}
Powerful active galactic nuclei, where particles can be accelerated up
to extreme energies, are often considered as the best astrophysically
motivated sources of energetic cosmic rays. The significant
correlations (${\cal{P}}=5\cdot10^{-3}$) of the highest-energy (at least one
standard deviation above $8\cdot10^{19}$~eV) events observed by Fly's
Eye, AGASA and Haverah Park and published before 1998 with flat-spectrum
radio-loud quasars (FSRQs) were first claimed in Ref.~\cite{Farrar} and
considered as an argument in favour of GZK-violating new composite
particles. Since then, these correlations were reanalyzed in various
papers \cite{...,Sigl} with different results. The original sample
consisted of quasars from the Bonn catalog \cite{Kuehr} of radio sources
with 5~GHz flux exceeding 1~Jy and radio spectral index $\alpha>-0.5$. We
considered both the original sample and the sample selected by the same
criteria from the new 11th edition of the V\'eron catalog of quasars
\cite{Veron2003}.

\paragraph{Catalog 1: Kuehr FSRQs} ~

{\bf Original catalog:} Kuehr et.\ al., 1981 \cite{Kuehr}.

{\bf Cuts:} class$=$QSO, spectral index $\alpha>-0.5$.

{\bf Number of objects:} 201.

\paragraph{Catalog 2: V\'eron FSRQs} ~

{\bf Original catalog:}
V\'eron-Cetty and V\'eron 2003 \cite{Veron2003}, Table
1 (quasars).

{\bf Cuts:} 6 cm flux $F_6>1$~Jy, spectral index $\alpha>-0.5$ (calculated
from fluxes at 6 and 11~cm).

{\bf Number of objects:} 211.

\subsection{Seyfert galaxies}
This subclass of active galaxies was considered in Refs.~\cite{Uryson}
where physical motivations were discussed and correlations (with the
probability of chance coincidence as low as $2\cdot 10^{-5}$) with
AGASA and Yakutsk cosmic rays with energies $E>4 \cdot 10^{19}$ were
claimed for several samples with fine-tuned cuts in the galactic
latitude.  In Ref.~\cite{Uryson}, the catalog \cite{Lipovetsky} and
the 10th edition of V\'eron catalog \cite{Veron2001} were used and the
distance cut $z<0.01$ was imposed. We use both the
catalog~\cite{Lipovetsky} and the updated V\'eron
catalog~\cite{Veron2003} with the same cut.

\paragraph{Catalog 3: Lipovetsky Seyferts} ~

{\bf Original catalog:}  Lipovetsky et. al. \cite{Lipovetsky}.

{\bf Cuts:} $z<0.01$.

{\bf Number of objects:} 136

\paragraph{Catalog 4: V\'eron Seyferts} ~

{\bf Original catalog:}
V\'eron-Cetty and V\'eron 2003 \cite{Veron2003}, Table 3 (active galaxies)

{\bf Cuts:} classification starts with S (all Seyferts), $z<0.01$.

{\bf Number of objects:} 190.

\subsection{BL Lac type objects}
Active galaxies often have jets fueled by the nuclei and consisting of
relativistic plasma fired in strongly collimated cones. When one of the
jets is directed towards the observer, this galaxy is seen as a blazar.
These objects are naturally considered as good candidates for the sources
of the cosmic rays which reached the Earth. BL Lac type objects constitute
a subclass of blazars distinguished by their spectral properties.
Different authors use different criteria for this distinction. In
particular, one may either require the absence of the emission lines or put
a constraint on their width. These criteria are discussed, for instance,
in Ref.~\cite{Veron-review} and are implemented in the
catalog~\cite{Veron2003} where Table 2 contains both confirmed BL Lac's in
the strict sense of the word and possible BL Lac's, notably those
classified as high-polarization (HP) blazars.

The first claim for correlation of cosmic rays
with confirmed BL Lac's (${\cal{P}}\sim 6\cdot 10^{-5}$)  was made
\cite{BL1} for the maximally autocorrelated sample of cosmic rays detected
by the AGASA and Yakutsk experiments. The sample of BL Lacs was chosen
from the 9th edition of the V\'eron catalog~\cite{Veron2000} by the cuts
on the visual magnitude, the radio flux and the redshift. The sample was
chosen in such a way that the signal was maximized; cited probability
includes the penalty factor for cut adjustment~\cite{BL1}.

As discussed below (see Sec.~3.8), gamma-ray emission should accompany
emission and propagation of UHECR. It was observed indeed that the BLL
which correlate with UHECR in the resulting subsample of Ref.~\cite{BL1}
are classified as EGRET sources in astronomical databases. Imposing the
single criterion -- positional coincidences of objects in
V\'eron~\cite{Veron2001} and EGRET~\cite{3EG} catalogues -- leaves 14
objects and this subsample also exhibits very low formal values of
${\cal{P}}\sim 3\cdot 10^{-7}$ with the maximally autocorrelated set of
UHECR~\cite{BL:EGRET}. It is difficult to calculate the significance here
since the set of 14 BLL overlaps with previous set of 22 BLL of
Ref.~\cite{BL1}. For our comparative study we have chosen one of these
subsamples (the other one shows very similar values of ${\cal{P}}$ in all
tests we carry). Namely, we have chosen single physical criterion of BLL
being the EGRET source for our catalog selection. This condition is more
transparent and is easier to implement with updated most recent catalogue
of AGN ~\cite{Veron2003}, which we use uniformly throughout our
comparative study.

A much more general sample of objects ---  all confirmed BL Lac's brighter
than $18^{\rm m}$ from the 10th edition~\cite{Veron2001} of the V\'eron
catalog --- was used, with a positive result, for the searches for
correlations with all published AGASA events (see
Refs.~\cite{BL:GMF,BL:subGZK,BL:KS}). In particular, it has been
noted~\cite{BL:GMF} that the account of the GMF improves the correlations.

Very recently, the correlations of all three previously selected
samples of the BL Lac type objects were tested~\cite{BL-HiRes} with
the HiRes stereo data with the positive result -- probability
${\cal{P}}=4\cdot 10^{-4}$ -- for the sample~\cite{BL:GMF,BL:subGZK,BL:KS} of
bright ($m<18$) confirmed BL Lac's.

The correlations with BL Lacs were reported also
in Ref.~\cite{Uryson1}.

For this study, we use the updated catalog~\cite{Veron2003} and do not
impose any new cuts. We consider separately "strict" and
high-polarization BL Lac's. We also consider the subsamples of the new
catalog~\cite{Veron2003} selected by the criteria of
Refs.~\cite{BL:GMF} and \cite{BL:EGRET}.
The results for the old~\cite{Veron2001,Veron2000} and
new~\cite{Veron2003} catalogs are similar.

\paragraph{Catalog 5: Confirmed V\'eron BL Lac's} ~

{\bf Original catalog:}
V\'eron-Cetty and V\'eron 2003 \cite{Veron2003}, Table
2 (BL Lac type objects).

{\bf Cuts:} classification$=$BL (confirmed BL Lac's).

{\bf Number of objects:} 491.

\paragraph{Catalog 6: HP V\'eron BL Lac's} ~

{\bf Original catalog:}
V\'eron-Cetty and V\'eron 2003 \cite{Veron2003}, Table
2 (BL Lac type objects).

{\bf Cuts:} classification$=$HP (high-polarization BL Lac's)

{\bf Number of objects:} 68.

\paragraph{Catalog 7: Bright confirmed V\'eron BL Lac's} ~

{\bf Original catalog:}
V\'eron-Cetty and V\'eron 2003 \cite{Veron2003}, Table
2 (BL Lac type objects).

{\bf Cuts:} classification$=$BL (confirmed BL Lac's), visual magnitude
$m<18$.

{\bf Number of objects:} 178.

\paragraph{Catalog 8: Possible EGRET BL Lac's} ~

{\bf Original catalog:}
V\'eron-Cetty and V\'eron 2003 \cite{Veron2003}, Table
2 (BL Lac type objects).

{\bf Cuts:} classification$=$BL (confirmed BL Lac), angular distance to the
nearest source from the 3EG catalog~\cite{3EG} less than $2r_{95}$, where
$r_{95}$ is the angular radius of the circle containing the same solid
angle as the 95$\%$ confidence contour of that EGRET source.

{\bf Number of objects:} 30.

\subsection{Radio galaxies}
Often intrinsically as luminous as blazars, similar objects seen at
large angles to the jet appear much fainter and are called radio
galaxies. Their classification has been introduced in Ref.~\cite{FR};
the FR~I type galaxies are thought to be the counterparts of the BL
Lac's while more powerful FR~II type objects may be the counterparts
of other blazars. The lobes (clouds of matter fueled by the jets) and
the hot spots (regions where the jet hits the lobe) of FR~II type
galaxies are efficient acceleration sites~\cite{Biermann}. According
to simple estimates (see, for instance, Ref.~\cite{AugerDesign}), the
hot spots are the only known places in the Universe where protons can
be accelerated to energies higher than $10^{20}$~eV by shock
acceleration. The shocks move in various directions, the magnetic
fields in the source are very strong and, as a result, the accelerated
particles may escape the accelerator not necessarily along the jet. We
are not aware of any previous correlation searches for the radio
galaxies. Here, we use a catalog of radio galaxies with their FR
classifications published in Ref.~\cite{FRcat}.

\paragraph{Catalog 9: FR~I radio galaxies} ~

{\bf Original catalog:} Zirbel and Baum \cite{FRcat}, Table 11.

{\bf Cuts:} classification$=$I or Ig (confirmed FR~I).

{\bf Number of objects:} 73.

\paragraph{Catalog 10: FR~II radio galaxies} ~

{\bf Original catalog:} Zirbel and Baum \cite{FRcat}, Table 11.

{\bf Cuts:} classification$=$II, IIg, IIn, or IId (confirmed FR~II).

{\bf Number of objects:} 116.

\subsection{Colliding galaxies}
Regions of strong shocks are obviously present in merging galaxy systems.
Based on this fact and on several positional correlations (without a
detailed statistical analysis),
nearby
colliding galaxies were suggested as
possible sources of the highest-energy cosmic rays
(see, for instance, Refs.~\cite{AGASA:list,AGASA:colliding}). Our sample
consists of objects classified as ``pair in contact'' in the latest update
of the Vorontsov-Velyaminov catalog of interacting
galaxies~\cite{VorontsovV}.

\paragraph{Catalog 11: Colliding galaxies} ~

{\bf Original catalog:} Vorontsov-Velyaminov et. al.  ~\cite{VorontsovV}.

{\bf Cuts:} classification starts with PK (pair in contact).

{\bf Number of objects:} 454.

\subsection{Starburst and luminous infrared galaxies}
In a region of large-scale star formation (a starburst region), a
strong shock wave may develop which provides the conditions required
for the efficient particle acceleration. Two-step acceleration process in
these regions may result~\cite{starbursts} in nuclei energies of
$10^{20}$~eV and higher.  When two colliding galaxies finally merge,
they may produce an active starburst region near the common center;
this appears to the observer as a luminous infrared galaxy (LIG).  The
possibility that the latter can be a source of highest-energy
cosmic rays has been discussed in Ref.~\cite{Smialkowski}. Given a
large number of sources, the correlation study was possible only in
the sense that the large-scale distribution of cosmic rays and LIGs
are correlated; these studies did not give definitive
results~\cite{Smialkowski,194}. In Ref.~\cite{Anch-review}, the most
probable candidates among LIGs were pointed out: those listed in the
PDS~\cite{PDS} and HCN~\cite{HCN} catalogs. It has been pointed out in
Ref.~\cite{Anch-review} that it is worth considering the list of
Ref.~\cite{Torres-LIG} in correlation studies. The latter list
consists of those LIGs for which the expected gamma-ray flux
(calculated in Ref.~\cite{Torres-LIG}) exceeds the sensitivity limit
of GLAST, a future gamma-ray telescope. Though it is not perfectly
clear why this particular GLAST-related cut is important for {\em
ultra-high-energy} cosmic ray production, we include the sample of
Ref.~\cite{Torres-LIG} in our study together with the PDS and HCN
samples.

\paragraph{Catalog 12: PDS starbursts} ~

{\bf Original catalog:} The Pico dos Dias Survey~\cite{PDS}, Table 1.

{\bf Cuts:} no.

{\bf Number of objects:} 203.

\paragraph{Catalog 13: HCN luminous infrared galaxies} ~

{\bf Original catalog:} HCN survey~\cite{HCN}, Table 1.

{\bf Cuts:} no.

{\bf Number of objects:} 53.

\paragraph{Catalog 14: Selected luminous galaxies} ~

{\bf Original catalog:} HCN~\cite{HCN} and PDS~\cite{PDS} surveys,
compiled in Ref.~\cite{Torres-LIG}.

{\bf Cuts:} calculated in Ref.~\cite{Torres-LIG} expected gamma-ray flux
exceeds the GLAST sensitivity.

{\bf Number of objects:} 16.

\subsection{Dead quasar candidates in luminous nearby galaxies.}
The observation of the cosmic rays with extremely high energies, at
which the propagation length of a particle is limited due to its
interaction with the cosmic background radiation, suggested that their
sources should be located in the neighborhood of our Galaxy. Since no
objects, powerful enough, are explicitly seen at these distances, it
has been proposed that the highest-energy cosmic rays may be
accelerated in quasar remnants which do not emit strong radiation
observed by astronomical, in particular radio,
telescopes~\cite{deadQSO,deadQSO1}.
A typical feature of a source of this kind is an extremely heavy
black hole in the nuclei of an elliptic galaxy. In Ref.~\cite{deadQSO}, a
list of possible candidate sources has been suggested; we included it in
our study. According to a refined study of Ref.~\cite{Torres}, a typical
dead quasar candidate which has chances to correlate with the observed
cosmic rays should satisfy the following criteria:

 1) absolute magnitude $M_{\rm abs}<-21$;

 2) redshift $z<0.01$;

 3) morphological type $t<-3$ (elliptical);

 4) galactic latitude $|b|>20^\circ$;

 5) the object belongs to a group with richness not exceeding 50.

The authors of Ref.~\cite{Torres} applied these five cuts to the
Nearby Optical Galaxy catalog~\cite{NOG}. The resulting sample
consisted of 12 objects. We did not succeed in reproducing their
catalog of 12 objects by the same procedure, probably due to the fact
that the cut (5) is quite vague and Ref.~\cite{Torres} used
unpublished data for its implementation. In our study, we skipped this
cut and left only first four. We applied them to the LEDA
database~\cite{LEDA} which is more complete than Ref.~\cite{NOG}. Still,
not all 12 objects of Ref.~\cite{Torres} entered
the resulting sample. Taking the original 12-object sample, we reproduced
the results for the AGASA experiment (without the GMF correction) presented
in Ref.~\cite{Torres}. This sample exhibits no correlations either upon
account of the GMF or with the HiRes data.

\paragraph{Catalog 15: dead QSO candidates from Ref.~\cite{deadQSO}} ~

{\bf Original catalog:}
sample of nearby galaxies with massive dark objects,
Ref.~\cite{Maggorian}.

{\bf Cuts:} central object mass $M>10^9 M_\odot$.

{\bf Number of objects:} 14.

\paragraph{Catalog 16: dead QSO candidates with cuts of
Ref.~\cite{Torres}} ~

{\bf Original catalog:} LEDA database~\cite{LEDA}.

{\bf Cuts:}  $M_{\rm abs}<-21$,  $z<0.01$,   $t<-3$, $|b|>20^\circ$.

{\bf Number of objects:} 23.

\subsection{Gamma ray sources}
Both acceleration of particles and their propagation through the cosmic
background radiation are accompanied by emission of energetic photons.
These photons, in turn, lose energy in interactions with background
photons~\cite{Berezinsky?} and arrive at the Earth with energies of
order 100~MeV -- 100~GeV, that is in the range observed by the EGRET
instrument. Hence, the gamma-ray emission is very probable to accompany the
cosmic ray emission (see, for instance,
Refs.~\cite{BL:EGRET,Blasi:gamma,Fargion:gamma}). The TeV emission has
been suggested as a characteristic feature of the AGNs which accelerate
protons up to highest energies but are quiet in other energy ranges,
analogues of the ``dead quasars''~\cite{deadTeV}.

Correlations of the blazars present among the {\em identified} sources in
the Third EGRET catalog~\cite{3EG} were studied in Ref.~\cite{Sigl} with a
negative result. Most of the EGRET blazars are strong sources of the
FSRQ or HP type (see Sec.~3.1, 3.3). In Ref.~\cite{BL:EGRET}, strong
correlations (${\cal{P}}\sim 10^{-4}$) were found between the sample of
AGASA and Yakutsk cosmic rays and {\em unidentified} EGRET sources. An
important ingredient for these correlations was the inclusion of the GMF
correction. Another correlating sample of EGRET sources studied in
Ref.~\cite{BL:EGRET} consisted of 14 sources which have a confirmed BL Lac
nearby (see Catalog 8 in Sec.~3.3).

Here, we separately study the samples of EGRET sources (i)~identified as
confirmed AGNs in the 3EG catalog
and (ii)~unidentified (outside the Galactic plane). We
also include a sample of GeV sources laying outside the Galactic plane and
a small sample of extragalactic TeV sources. For comparison, we present
the results for unidentified EGRET and GeV sources without the galactic
latitude cut.

\paragraph{Catalog 17: EGRET blazars} ~

{\bf Original catalog:}
The 3rd EGRET catalog ~\cite{3EG}.

{\bf Cuts:}  identification$=$A (confirmed AGN).

{\bf Number of objects:} 67.

\paragraph{Catalog 18: Unidentified EGRET sources outside the Galactic
plane} ~

{\bf Original catalog:}
The 3rd EGRET catalog ~\cite{3EG}.

{\bf Cuts:} identification void, galactic latitude $|b|>10^\circ$.

{\bf Number of objects:} 96

\paragraph{Catalog 18a: Unidentified EGRET sources} ~

{\bf Original catalog:}
The 3rd EGRET catalog ~\cite{3EG}.

{\bf Cuts:} identification void.

{\bf Number of objects:} 170

\paragraph{Catalog 19: GeV sources outside the Galactic
plane} ~

{\bf Original catalog:} Lamb, Macomb \cite{GEV}.

{\bf Cuts:} galactic latitude $|b|>10^\circ$.

{\bf Number of objects:} 26

\paragraph{Catalog 19a: GeV sources} ~

{\bf Original catalog:} Lamb, Macomb  \cite{GEV}.

{\bf Cuts:} no.

{\bf Number of objects:} 52.

\paragraph{Catalog 20: Extragalactic TeV sources} ~

{\bf Original catalog:} List of confirmed TeV sources presented in
Ref.~\cite{TeV:ICRC}.

{\bf Cuts:} identification with extragalactic objects\footnote{The only
TeV source present in the list of Ref.~\cite{TeV:ICRC} and lacking firm
identification is most probably Galactic and is not included.}.

{\bf Number of objects:} 10.

\section{Results}
For the catalogs described in Sec.~3, we estimated, as described in Sec.~2,
the probability that the observed number of cosmic-ray events close to the
astrophysical objects is a fluctuation of the random
distribution.

\subsection{Cosmic-ray samples}
The samples of cosmic rays we use are determined mostly by the
availability of data. Most of the previous studies were based on AGASA
and (sometimes) Yakutsk events. The HiRes stereo sample became
recently available. Motivated by the fact that several
experimental groups published only the data on the events with energies
higher than $10^{20}$~eV and keeping in mind the possibility that the
``sub-GZK'' and ``super-GZK'' events may originate in different classes of
objects, we consider separately the events with $E \gtrsim 10^{20}$~eV. We
describe the cosmic-ray sets in Table~1.
\begin{table}
\begin{center}
\begin{tabular}{|l|c|c|c|c|l|}
\hline
Sample & Experiments & Energies & Reference & Number of  & Notes\\
&&&&cosmic rays &\\
\hline
A & AGASA & $(4\dots 10)\cdot 10^{19}$~eV& \cite{AGASA:list} & 49 &
     \\
A$+$Y &AGASA,&$(4\dots 10)\cdot 10^{19}$~eV&\cite{AGASA:list}, &59 &
     \\
      & Yakutsk &                          &\cite{Pravdin} &      &1
     \\
H     & HiRes stereo& $>10^{19}$~eV& \cite{HiRes} & 271 & 2
     \\
$20+$ & All Northern& $>10^{20}$~eV& See Ref.~\cite{OurSample} & 13 &3
     \\
      & experiments &              &                           &     & \\
\hline
\end{tabular}
\end{center}
\caption{The cosmic-ray samples. {\em Notes:} 1)~The updated sample of
events observed by the Yakutsk array before 1995 with zenith angles
less than $45^\circ$.
2)~The arrival directions were published in the
form of a PS plot in Hammer projection; the energies of the events are
unpublished, so the correction for the GMF is not possible.
3)~The sample is described in detail in
Ref.~\cite{OurSample}.
}
\end{table}

\subsection{Results of the comparative study}
The values of the probability ${\cal P}$ calculated following Sec.~2 are
presented in Table~2.
\begin{table}
\begin{center}
\begin{tabular}{||c|c|c|c|c||c||c||}
\hline
Catalog & A & A & A$+$Y & A$+$Y & H & $20+$ \\
        &   & * &       &  *    &       &   \\
\hline
1 (FSRQ)&0.23&1.00&0.20&0.88&              0.69&0.73\\
2 (FSRQ)&0.95&0.98&0.96&0.92&              0.28&1.00\\
3 (Sy)&0.70&{\bf 0.013}&0.45&{\bf 0.038}&0.86&1.00\\
4 (Sy)&0.69&0.62&0.79&0.78&              0.40&0.74\\
5 (BL)&0.17&{\bf 0.024}
&0.29&{\bf 2.9$\bf \cdot 10^{-3}$}& {\bf 0.040}&0.98\\
6 (BL)&0.59&0.56&0.44&0.77&              0.078&0.48\\
7 (BL)&{\bf 0.027}&{\bf \underline{$\bf 4\cdot 10^{-4}$}}&
 0.051&{\bf \underline{$\bf 7\cdot 10^{-4}$}}
& {\bf \underline{$\bf 5\cdot 10^{-4}$}}& 0.87\\
8 (BL)&{\bf 7.4$\cdot \bf 10^{-3}$}&0.20&{\bf 2.9$\bf \cdot
10^{-3}$} &{\bf 0.027}&0.54&1.0\\
9 (RG)&0.56&0.56&0.58&0.46   &1.00&0.49\\
10 (RG)&0.89&0.34&0.95&0.48  &0.63&0.68\\
11 (coll.)&0.27&0.15&0.069&0.099&0.55&0.50\\
12 (LIG)&0.089&0.47&0.18&0.27 &0.80&0.22\\
13 (LIG)&0.11&0.87&0.18&0.11  &0.30&0.41\\
14 (LIG)&0.086&0.18&0.13&0.090&0.24&1.00\\
15 (dQSO)&1.00&0.17&1.00&0.22& 1.00 & 1.00  \\
16 (dQSO)&0.41&1.00&0.50&1.00  &1.00&0.13\\
17 ($\gamma$) &0.24&0.68&0.14&0.80  &0.20&0.46\\
18 ($\gamma$)&0.10&{\bf 5.1$\bf \cdot 10^{-3}$}&0.16&{\bf \underline{1.2$\bf \cdot
10^{-3}$}}&0.62&1.00\\
18a ($\gamma$)&0.24&{\bf 0.015}&0.37&{\bf 5.8$\bf \cdot 10^{-3}$}&0.82&0.61\\
19 ($\gamma$)&0.27&0.31&0.34&1.00&0.40&1.00\\
19a ($\gamma$)&0.19&0.39&0.095&0.50&0.52&1.00\\
20 ($\gamma$)&1.00&1.00&1.00&1.00&{\bf 0.012}&1.00 \\
\hline
\end{tabular}
\end{center}
\caption{Probability ${\cal P}$ calculated for the catalogs of
astrophysical objects described in Sec.~3 and the samples of cosmic rays
described in Sec.~4.1.
For convenience, with the number of a catalog, a reference to a section of
this paper is given as follows: FSRQ -- Sec.~3.1, Sy -- Seyfert galaxies
-- Sec.~3.2, BL -- BL Lac's -- Sec.~3.3, RG -- radio galaxies -- Sec.~3.4,
coll.\ -- colliding galaxies -- Sec.~3.5, LIG -- starburst and luminous
IR galaxies -- Sec.~3.6, dQSO -- dead QSO candidates -- Sec.~3.7,
$\gamma$ -- gamma-ray sources -- Sec.~3.8.
The columns with the results obtained with
correction of the arrival directions for GMF are marked by
asterisks. The values of ${\cal P}<4.6\cdot 10^{-2}$ (which would
correspond to a $2\sigma $ effect for the Gaussian statistics) are given
in bold face. The values ${\cal P}<2.7\cdot 10^{-3}$ (would-be
$3\sigma $) are underlined.  Each value of ${\cal P}$ has been
calculated without accounting for other tries and thus can be
interpreted as a relative probability with respect to other
catalogues.}
\end{table}
One can make following conclusions out of these Tables.
\begin{enumerate}
\item
At present, cosmic rays with $E>10^{20}$~eV do not correlate with any
astrophysical objects. This does not exclude, however, any real connection
because of low statistics.
\item
The strongest correlations seen in AGASA and Yakutsk datasets at
energies below $10^{20}$~eV are exhibited by the catalogs 3, 5, 7, 8 and
18.
\begin{enumerate}
\item
The effect in Catalog 3 (Lipovetsky Seyferts) is dominated by 5 sources
close to the direction of the AGASA triplet. These sources contribute 8
out of 13 observed coincidences with AGASA cosmic rays within
$2.8^\circ$. The effect is not present for a sample of Seyfert galaxies
from the V\`eron catalog. The correlations are observed with the account
of the correction for GMF only.
\item
The correlations of Catalog 5 (all confirmed BL Lac's) and Catalog 7
(bright confirmed BL Lac's) with AGASA and Yakutsk cosmic rays are improved
with the correction for GMF.
\item
The Catalog 8 represents an update of the sample of Ref.~\cite{BL:EGRET}
of BL Lac's which are possibly associated with EGRET sources. It contains
30 objects, compared to 14 original. The values of probabilities presented
here are higher compared to those given in
Ref.~\cite{BL:EGRET}. The difference is due to the fact that for the
comparative study, we use here different sets of cosmic rays. With the
update 30-objects catalog and the same maximally autocorrelated sample of
cosmic rays as used in Ref.~\cite{BL:EGRET}, we obtain the results very
similar to those presented there for the 14-object sample.
\item
Strong correlations of cosmic rays observed by the ground arrays are seen
with unidentified EGRET sources (catalog 18) outside the Galactic
plane. Generally, correlated sources were detected neither in GeV nor
in TeV.
The physical properties of the correlated
sources and their identification require further study. The
correlations with unidentified EGRET sources persist even when a large
Galactic population is included (Catalog 18a).  Note in passing
that on the base of the correlations with 14 BL Lac objects associated
with EGRET sources it was suggested~\cite{BL:EGRET} that at least some
of the unidentified EGRET sources, correlated with UHECR, are BL Lac
type objects still unobserved in other wave bands.

\end{enumerate}
\item
Results on AGASA and (AGASA$+$Yakutsk) samples are consistent with each
other.
\item
The corrections for the deflection in GMF are important. A full test of the
correlations found in the AGASA and Yakutsk data with the new HiRes stereo
data will be possible only after publication of the energies of HiRes
events. One should note however that the poor knowledge of the GMF results
in errors in corrected arrival directions which exceed the angular
resolution of the HiRes experiment in the stereo mode, so the effect, if
present, may be smoothed considerably.
\item
Only BL Lac type objects (catalogs 5, 7) exhibit correlations both with the
HiRes stereo data set and the AGASA-dominated samples. This fact is very
interesting and deserves further discussion; see also
Sec.~\ref{sec:tries}.
To the correlations of HiRes cosmic rays with the TeV sources (catalog
20), two objects contribute, and both are BL Lacs. Other catalogs are not,
however, rejected. For instance, the correlations with unidentified
gamma-ray sources are mostly seen for the proton primaries when the
deflection of arrival directions in GMF were taken into account;
a {similar analysis} for the HiRes sample is impossible at the moment
because the energies of the cosmic rays are unpublished. Moreover, the
angular resolution of HiRes stereo is better than uncertainties both in
the positions of the EGRET sources (which means that the ultra-high-energy
astronomy is actually possible) and in the corrections for GMF.
\end{enumerate}

\subsection{Significance of multiple tries}
\label{sec:tries}
When the purpose of a study is to estimate the probability that some
particular class of objects is related to a sort of events (UHECRs in our
case), multiple tries of various catalogs are not welcome. Just by chance,
one expects the probability of order $1/M$ to be found among $M$
independent catalogs of objects tested for correlations with a single
cosmic-ray sample. Thus, a factor of $M$ is usually required as a penalty
in this case. When the tries are not independent, the penalty factor is
more difficult to estimate. The purpose of our study is, however, purely
comparative: we test the conjectures made in previous works by a uniform
method. To compare the probabilities, we thus do not care about any general
factor by which all the numbers should be multiplied, and this penalty
factor in particular.

On the other hand, one notices that quite low ${\cal P}$ are found for
the Catalog 7 with the cosmic rays observed by different
experiments. While this is not our main goal, one may wonder how
probable is to obtain ${\cal P}\le 7 \cdot 10^{-4}$ in {\it
independent} experiments just by chance.  In this way, one would test
the hypothesis that there are no catalogs correlating with UHECR
samples among all we list and all the correlations seen are nothing but
statistical fluctuations due to multiple tries. We performed a simple
estimative simulation to this end. We took the declinations of objects
present in our 21 catalogs and randomized the right ascensions many times.
In this way, we obtained $32000$ sets, each of 21 catalog with the same
numbers of objects and the same distributions in declination as the real
ones. All $21\cdot 32000\approx 7\cdot10^5$ simulated catalogs were
subject to the same procedure as the real catalogs. Only in one case, a
simulated catalog exhibited ${\cal P}\le 7 \cdot 10^{-4}$ for the HiRes
sample and one of four other (interdependent) samples simultaneously. We
thus estimate the probability that the results for the Catalog 7 are a
result of a chance fluctuation as $1/32000\approx 3\cdot10^{-5}$.

\subsection{Dependence on $\delta $}

As it has been already noted, the values of the bin sizes we fixed for
the calculation of ${\cal P}$ were determined by simulations. In some
cases, lower values of ${\cal P}$ were obtained for other choices of
$\delta $. In figures 1 and 2,
\begin{figure}
\centerline{\includegraphics[width=142mm]{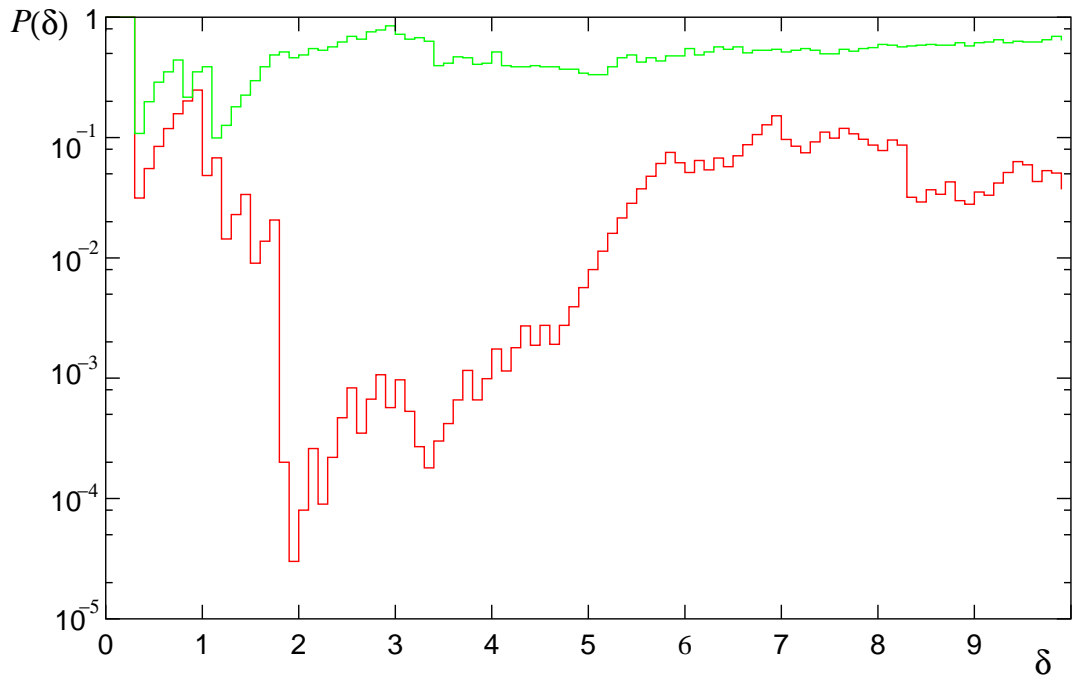}}
\caption{${\cal P}(\delta )$ for catalog 18 (unidentified EGRET sources
outside the Galactic plane, lower (red) curve) and catalog 4 (nearby
V\'eron Seyferts, upper (green) curve), cosmic-ray sample A$+$Y (AGASA and
Yakutsk), correction for GMF included. }
\end{figure}
\begin{figure}
\centerline{\includegraphics[width=142mm]{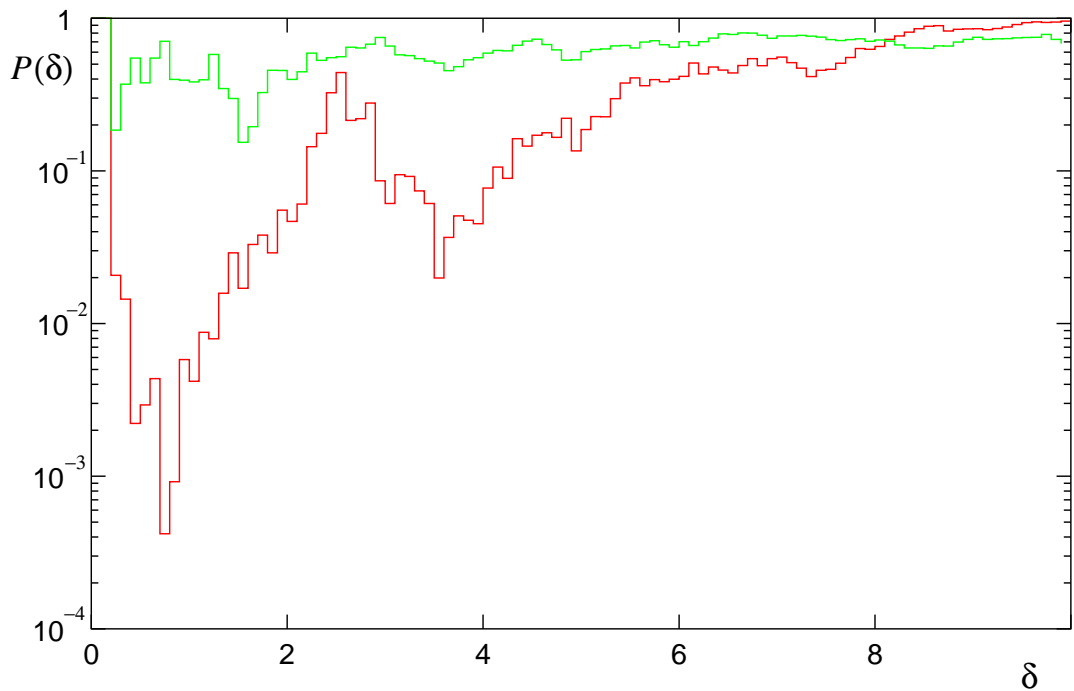}}
\caption{${\cal P}(\delta )$ for catalog 7 (bright confirmed BL Lac's,
lower (red) curve) and catalog 4 (nearby V\'eron Seyferts, upper (green)
curve), cosmic-ray sample H (HiRes stereo). }
\end{figure}
we present the $\delta $ dependence of ${\cal P}$ for
two interesting cases: unidentified EGRET sources with AGASA and
Yakutsk cosmic rays and BL Lac's with HiRes cosmic rays. For comparison,
we present on the same plots the ${\cal P}(\delta )$ dependence for a
typical catalog which does not exhibit any correlation. An example of the
effect described in Sec.~2 -- a random coincidence at very small angles --
is presented at Fig.~3.
\begin{figure}
\centerline{\includegraphics[width=142mm]{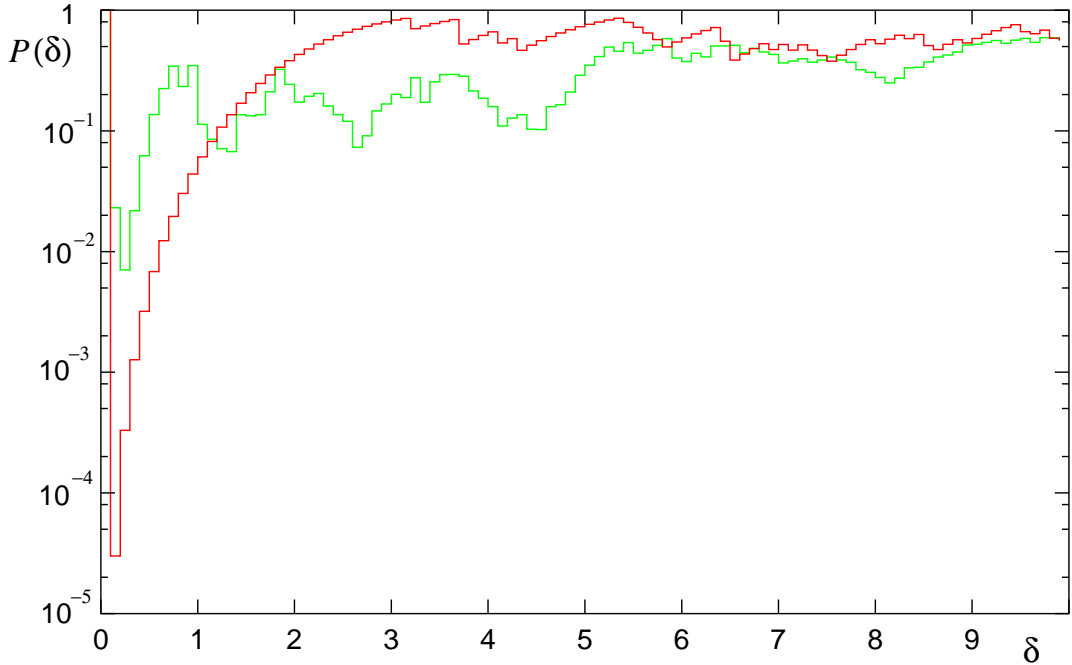}}
\caption{Curve with a shallow minimum (green): ${\cal P}(\delta )$ for
catalog 9 (colliding galaxies), sample A, GMF included. The dip at $\delta
=0.3^\circ$ is probably unphysical and diminishes for $\delta =2.8^\circ$
expected from the experiment angular resolution. Curve with deeper minimum
(red): catalog 20 (extragalactic TeV sources), sample H. For this sample,
${\cal P}(0.2^\circ)< 10^{-4}$ while physically meaningful ${\cal
P}(0.8^\circ)\sim 10^{-2}$.}
\end{figure}

\section{Conclusions}
We performed a comparative study of correlations between the arrival
directions of the cosmic rays observed by AGASA, Yakutsk and HiRes
experiments with 21 catalogs of astronomical objects suggested by
different authors as possible UHECR sources. No a posteriori cuts were
imposed on the catalogs. The results are described in Sec.~4.2 but should
be interpreted with great care because the angular resolution of the
cosmic-ray experiments is poor, cosmic magnetic fields are known not quite
well and the statistics is not overwhelming.

The analysis of the AGASA and Yakutsk samples demonstrates the presence of
correlations with unidentified EGRET sources, BL Lac's and nearby Seyfert
galaxies.
The corrections for the Galactic magnetic field are very important for
these effects.

Only BL Lac type objects exhibit correlations with the HiRes stereo
sample as well. With the uncut catalog of 491 confirmed BL Lac's and the
set of 271 cosmic rays with energies $E>10^{19}$~eV, we obtained the
probability of $0.04$ for these correlations to result from a random
coincidence. For the updated catalog of 178 bright ($m<18$) BL Lac's, the
probability is as low as $5\cdot 10^{-4}$. For the AGASA sample (with the
GMF correction), ${\cal P}=4\cdot 10^{-4}$ for the same catalog; for AGASA
and Yakutsk ${\cal P}=7\cdot 10^{-4}$. The probability that one of 21
catalogs exhibits, by chance, ${\cal P}\le 7\cdot 10^{-4}$ both for the
HiRes sample and one of four other samples is about $3\cdot 10^{-5}$. The
correlations are present at the angular distances of order HiRes angular
resolution and much less than the expected deflections of charged
particles in cosmic magnetic fields. This suggests that a fraction of the
cosmic-ray flux at energies $E>10^{19}$~eV is carried by neutral
primaries~\cite{BL-HiRes}.

Clearly, more detailed information about the cosmic rays observed by
HiRes would be very important for understanding the effects discussed
here.  Knowledge of the energies of particles would allow to
reconstruct (to some accuracy) their deflections in the Galactic
magnetic field. A study of shower profiles might help to understand
what kind of neutral particles may be present in the cosmic rays and
contribute to correlations. Finally, if, for example, a neutral
primary is a photon emitted by a blazar, the arrival time of the
corresponding air shower may coincide with a temporary change
in the state of the source.  This would give much stronger support to
conjectures about particular cosmic-ray sources than the statistical
studies of the arrival directions.

Further detailed study will specify which particular
physically distinguished subclasses of BL Lac's and other objects
correlate strongly with the cosmic rays and will give quantitative
predictions for the future experiments, in particular for the Pierre Auger
Observatory.

\section*{Acknowledgements}
The authors are indebted to P.~Tinyakov who shared with them his code
which was used to calculate the probabilities with the correction for GMF
and to check other results for the ground arrays. We thank D.~Semikoz,
P.~Tinyakov and I.~Tkachev for numerous helpful discussions and
reading of the manuscript.
We thank
M.~Pravdin and I.~Sleptsov for sharing with us the updated catalog of
Yakutsk events.
S.T.\ thanks {\em Service de Physique Th\'eorique}, {\em Universit\'e Libre
 de Bruxelles}, for warm hospitality at different stages of the work,
 supported in part by the IISN (Belgium), the {\em Communaut\'e
Fran\c{c}aise de Belgique} (ARC), and the Belgium Federal Government
(IUAP). This study was supported in part by the INTAS grant 03-51-5112, by
the Young scientists of the Russian Academy of Sciences Project \# 35, by
the RFBR grant 02-02-17398, by the grants of the President of the Russian
Federation NS-2184.2003.2, MK-2788.2003.02 (D.G.), MK-1084.2003.02 (S.T.),
by the Swiss Science
Foundation, grant 20-67958.02,
by the grants of the Russian Science Support Foundation and by the
fellowships of the "Dynasty" foundation (awarded by the Scientific Council
of ICFPM).

\end{document}